\documentclass[doublecol]{epl2}

\title{A robust ranking algorithm to spamming}
\author{Yan-Bo Zhou\inst{1} \and Ting Lei\inst{1} \and Tao Zhou\inst{2}\thanks{E-mail: \email{zhutou@ustc.edu}}}
\shortauthor{Yan-Bo Zhou \etal}
\institute{
  \inst{1} D\'{e}partement de Physique,  Universit\'{e} of Fribourg - CH-1700 Fribourg, Switzerland\\
  \inst{2} Web Sciences Center,  University of Electronic Science and Technology of China - 610054 Chengdu, PRC
}


\abstract{
Ranking problem of web-based rating system has attracted many attentions. A good ranking algorithm should
be robust against spammer attack. Here we proposed
a correlation based reputation algorithm to solve the ranking problem of such rating systems where
user votes some objects with ratings. In this algorithm, reputation of user is iteratively determined by the
correlation coefficient between his/her rating vector and the corresponding objects' weighted average rating vector.
Comparing with iterative refinement (IR) and mean score algorithm, results for both artificial
and real data indicate that, the present algorithm shows a higher robustness against spammer attack.
}

\begin{document}

\maketitle

\section{Introduction}

The abundance of available information troubled people every day, and information filtering technique is quickly
developed in recent years. An important aspect in information filtering is the rating system. There are a range of daily examples
of rating system. Such systems include opinion websites (Ebay, Amazone, Movielens, Netflix, etc.),
where users evaluate objects. Ranking is one of the most common way to describe the
evaluation aggregation result, which gives a simple representation of the comparative qualities of objects.

PageRank is the most widely applied algorithm for search engines which rank websites 
based on the directed hyperlink graph \cite{PR}. Recently, some iterative algorithms are used in scientific 
citation network to rank scientists \cite{scientist}. Both the hyperlink network and scientific citation network are
unipartite systems, but many other rating systems have a bipartite structure
with two kinds of node: users as evaluators and objects as candidates \cite{distribution}. In this paper, we 
consider the ranking problem in such rating systems where users vote objects with ratings, and
devise algorithms to accurately rate objects.

Ranking objects according to their average ratings is a straightforward statistical method. 
However, in the open evaluation system, the user can be somebody who are not serious about 
voting, or he/she is not experienced in the corresponding field and gives some unreasonable ratings. What
even worse is that the user might be an evil spammer who gives biased ratings on purpose. Therefore,
the evaluation by simply averaging all ratings may be less accurate. Building a reputation system
for users is a good way to solve this problem \cite{reputation1,reputation2}. Users with higher reputations are assigned more weight. Such
reputation mechanisms are widely used in online systems, such as online shops \cite{reputationshop},
online auctions \cite{reputationauction},
 Wikipedia \cite{reputationwiki}, P2P sharing networks \cite{reputationp2p}, etc.

There are already some ranking algorithms based on reputation estimate \cite{IR1,IR2,jiang,ker}. In \cite{IR1,IR2}, an iterative refinement
(IR) algorithm is proposed. A user's reputation is inversely proportional to the difference between his/her rating vector and the
corresponding objects' weighted average rating vector. Weighted rating of all objects and reputation of all users 
are recalculated at each step, until the change of weighted ratings is less than a certain threshold between
two iteration steps. Kerchove and Dooren \cite{ker} modify the iterative refinement algorithm by assigning trust to each
individual rating. In most previous works, the influence of spammer attack in rating
systems is always ignored.

In this paper, we proposed a correlation based ranking algorithm. Reputation of user is determined by the
correlation coefficient between the user's rating vector and the corresponding objects' weighted average rating vector.
By comparing with other algorithms, the effectiveness of the correlation based ranking algorithm was tested using 
artificial data. The results show that correlation based ranking algorithm is more robust than other algorithms.
Finally, we use two distinct real data sets (Movielens and Netflix) to evaluate the effectiveness of the algorithm.

\section{The correlation model}
The rating system we considered can be represented by a bipartite network, which consists of a set $U$
of users who have each rated some subset of the complete set $O$ of objects. We use Latin letters for users and Greek letters for objects to distinguish them.
Consequently $r_{i\alpha}$ denotes the rating given by user $i$ to object $\alpha$.
The set of users who rated a given object $\alpha$ is denoted by $U_\alpha$, while the set of objects rated by a user $i$ is denoted by $O_i$.
The degree of object $\alpha$ (i.e. the number of ratings given to object $\alpha$) is denoted as $ko_{\alpha}$ and the
degree of user $i$ (i.e. the number of ratings given by user $i$) is denoted as $ku_i$.

We use $Qo_{\alpha}$ to represent the aggregate estimated quality of object $\alpha$, and
$Cu_{i}$ the reputation of user $i$.
The quality of an object depends on the evaluations it received, and can be defined as the weighted average of ratings to
this object:

\begin{equation}
\label{eq.1}
Qo_\alpha=\frac{\sum_{i\in U_\alpha}{Cu_ir_{i\alpha}}}{\sum_{i \in U_\alpha}{Cu_i}}.
\end{equation}

According to the objects' qualities, the Pearson correlation coefficient between the rating vector of user $i$ and the corresponding objects' quality
vector is given by:
\begin{equation}
\label{eq.2}
Corr_i=\frac{1}{ku_i}\sum_{\alpha \in O_i}{(\frac{r_{i\alpha}-\overline{r_i}}{\sigma _{r_i}})(\frac{Qo_{\alpha}-\overline{Qo_i}}{\sigma _{Qo_i}})},
\end{equation}
where $\sigma _{r_i}$ and $\sigma _{Qo_i}$ are resoectively the standard deviations of rating vector of user $i$  and the corresponding objects' quality
vector, and $\overline{r_i}$ and $\overline{Qo_i}$ are their expected values.

Correlation coefficient is a good way to quantify the similarity between two vectors. As a user who has more similar ratings to the weighted average ratings should
have a higher reputation, the reputation of a user $i$ is given based on this similarity:

\begin{equation}
\label{eq.3}
Cu_i=\left\{
\begin{array}{rcl}
 corr_i & & \mbox{if $corr_i \ge 0$}\\
 0 & & \mbox{if $corr_i < 0$}
\end{array}
\right.
\end{equation}

The resultant object quality is obtained by initially assigning every user's reputation according to his degree as $Cu_i=ku_i/|O|$, 
and then iterating eq.~(\ref{eq.1}, (\ref{eq.2}) and (\ref{eq.3}) until
the change of the quality estimates
\begin{equation}
|Qo-Qo^{'}|=1/|O|\sum_{\alpha \in O}{(Qo_{\alpha}-Qo^{'}_{\alpha})^{2}},
\end{equation}
is less than a threshold of $\delta = 10^{-6}$.

\section{Results on artificial data}
When creating the artificial data, we assume that each user $i$ has a certain magnitude of rating error
$\delta_{i} (i = 1, . . . , |U| )$ and each object $\alpha$ has a certain true intrinsic quality $Q_{\alpha} (\alpha = 1, . . . , |O|)$.
At each time step $t$, a user-object pair $(i,\alpha)$, on which the rating $r_{i\alpha}$ has not been given (at all $t^{'}< t$), is chosen.
The rating $r_{i\alpha}$ is determined as
\begin{equation}
r_{i\alpha} = Q_{\alpha} + e_{i\alpha},
\end{equation}
where error $e_{i\alpha}$ is drawn from a probability distribution parameterized by user $i$'s error magnitude. Rating $r_{i\alpha}$
lying out of the range are truncated. To
achieve a certain sparsity $\eta$ of the resulting data, the total number of generated ratings
is $\eta |U||O|$ hence ($t = 1, . . . , \eta |U||O|$).

As evident from the power-law-like distribution of the
number of ratings given by individual users and received by individual objects in the real data sets \cite{distribution}, there should be a preferential
attachment mechanism in the evolution of the rating system \cite{BA}. In the real data sets, the more ratings a user have given, the higher
probability he will give a new rating. And the more ratings an object have received, the higher
probability it will receive a new rating. Based on these observations, at each time step $t$, we choose a
user-object pair $(i, \alpha)$ using preferential attachment mechanism. The probabilities of choosing user $i$ and object $\alpha$ at time step $t$ are
\begin{equation}
p_i(t)=\frac{ku_{i}(t)+1}{\sum_{j \in U}{(ku_{j}(t)+1)}}
\end{equation}
and
\begin{equation}
p_{\alpha}(t)=\frac{ko_{\alpha}(t)+1}{\sum_{\beta \in O}{(ko_{\beta}(t)+1)}},
\end{equation}
where $ku_{i}(t)$ and $ko_{\alpha}(t)$ are the degree of user $i$ and object $\alpha$ at time step $t$. As the degrees are
all zero at the initial time, we have used $ku_{i}(t)+1$ in the above equations.

To create artificial data, we set $|U| = 6 000$, $|O| = 4 000$ and $\eta = 2\%$ (which corresponds to approximately $4.8 \times 10^5$
ratings). Objects' qualities and users' ratings are limited to the range $[0, 1]$. Objects' qualities are drawn from the uniform distribution $U(0, 1)$, users' error magnitudes 
are drawn from the uniform distribution $U(\sigma_{min} , \sigma_{max} )$, and individual rating errors $e_{i\alpha}$ are drawn 
from the normal distribution $N(0, \sigma_{i})$. We choose $\sigma_{min}=0.1$ and $\sigma_{max}=0.5$ in the simulation.

\begin{figure}[h!]
\centering
\scalebox{0.5}[0.5]{\includegraphics{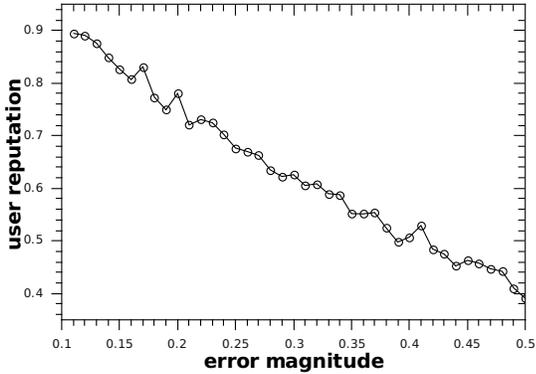} }\caption{
The relationship of user's reputation and his error magnitude of the correlation based ranking algorithm.}
\label{fig1}
\end{figure}

To get a more accurate ranking, a good ranking algorithm should give higher reputations to the users with lower error magnitudes.
As the users' error magnitudes are continuous, we divide the error magnitude into bins with the
length 0.01. The mean reputations of users with error magnitudes in the same bins are then evaluated. Fig.~\ref{fig1} shows the users' mean
reputation as a function of error magnitude obtained by the correlation based ranking algorithm. It is clear that the higher the error
magnitude of the user, the lower the reputation. The correlation coefficient is thus a good way to quantify a user's reputation.

After the convergence of $Qo$, we use a correlation measure called Kendall's tau \cite{tau} to judge the ranking result of
the algorithm. It is defined as
\begin{equation}
 \tau=\frac{2}{|O|(|O|-1)}\sum_{\alpha < \beta}{\mbox{sign}[(Q_{\alpha}-Q_{\beta})(Qo_{\alpha}-Qo_{\beta})]},
\end{equation}
with the lower bound -1 ($i.e.$ the two rankings are exactly opposite) and the upper bound 1 ($i.e.$ the two rankings are exactly
the same).

Besides, there is another standard measure in information filtering literature named AUC \cite{AUC}. In most cases, the
true ranking of objects is not available, and it is not possible to evaluate the algorithm by $\tau$. Instead,
we can select a group of benchmark objects by some plausible criteria, and then use AUC to evaluate a ranking algorithm.
AUC equals one when all benchmark objects are ranked higher than the other objects, while AUC=0.5 corresponds
a completely random ranked object list. In the tests using artificial data, 5\% of all objects with the highest quality values
are selected as benchmark objects.

Using the artificial data, we evaluate the effectiveness of the correlation based ranking algorithm. Comparing with straightforward
mean algorithm and IR algorithm, table~\ref{tab1} shows the ranking result obtained from the artificial data.
As we can see,
in a clean rating system without any spammer, the effectiveness of the three algorithms are all good and do not differ a lot.
The IR algorithm relatively has the best effectiveness.

\renewcommand\arraystretch{1.6}
\begin{table}
\caption{Ranking results of different algorithms
for the artificial data. }
\label{tab1}
\setlength\tabcolsep{9pt}
\begin{tabular}{cccc}

\hline
Algorithm & Mean & IR & Correlation-Based \\
\hline
\centering
AUC & 0.9940 & 0.9965 & 0.9952  \\		
$\tau$ & 0.9216 & 0.9387 & 0.9300 \\		
\hline
\end{tabular}
\end{table}

\section{Spam analysis}

In the above simulations, users are honest and give ratings with fixed error magnitudes. While in the real system, not all users are
honest. There are many kinds of spammers that may drastically lower the effectiveness of ranking algorithms.

In general, there are two kinds of ratings that a spammer may give: (1) Random rating: random allowable ratings on items.
(2) Push rating: maximum or minimum allowable ratings on items.

A random rating spammer may be a naughty user who just plays around with the informations and gives ratings which mean nothing. A push rating spammer always gives maximum/minimum
allowable ratings that also mean nothing. These dishonest ratings influence the accuracy of the ranking result. A good ranking algorithm should be robust
against any kind of spammers. To evaluate the correlation based ranking algorithm against different types of spammers, some users
are randomly selected as spammers in the artificial data. These spammers' ratings are generated according to their spamming types.
In this paper, we consider two types of spamming:
(1) Spammers who always give random ratings. (2) Spammers who always give push ratings. For both types of spamming, we study the influence on the effectiveness of the correlation based
ranking algorithm as the ratio of spammers increases. For comparison, the effectiveness of mean and IR ranking algorithm are also studied.

\subsection{Random rating spamming}

\begin{figure}[h!]
\centering
\scalebox{0.5}[0.5]{\includegraphics{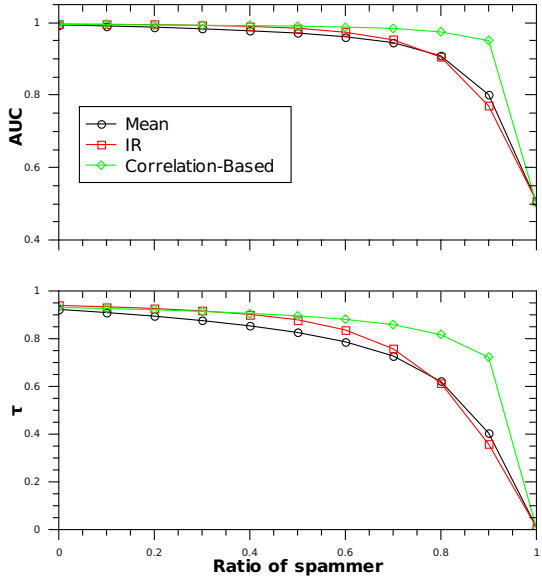} }\caption{
The effectiveness of different algorithms to random rating spamming. The result is obtained by averaging over 10 independently run.}
\label{fig2}
\end{figure}

Fig.~\ref{fig2} shows the effectiveness of different algorithms obtained from the artificial data with random rating spamming. When there is no spammer,
the effectiveness of all the three algorithms are almost the same. But when the ratio of spammers increases, the correlation based ranking algorithm is significantly better than the others.
When all the users are spammers, the rankings are random for all algorithm, and the value of AUC becomes 0.5 and $\tau$ becomes 0.

\begin{figure}[h!]
\centering
\scalebox{0.5}[0.5]{\includegraphics{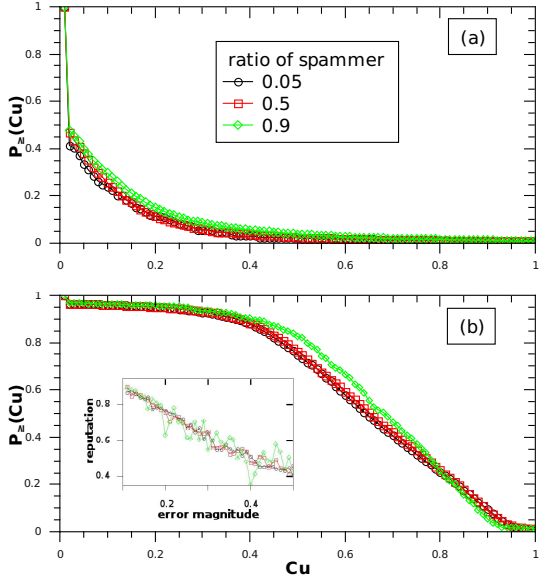} }\caption{
The distributions of reputations of (a) spammers  and (b) honest users with different ratio of spammers. The inset denote
the relationship of honest user's reputation and his error magnitude. All the spammers are
random rating spammers.}
\label{fig3}
\end{figure}

Correlation coefficient is a measure of the strength of the linear relationship between two vectors. A random value vector normally has little or no correlation
with any other vectors. Thus, the reputation of random rating spammers should be very small. As shown in fig.~\ref{fig3}(a), the reputations of most random rating spammers are very low.
 Even when the ratio of spammers is 0.9, there is still more than 70\% of spammers with the reputation less than 0.1.

While for the honest users, regardless of spammer size, their reputations are
always high (up to 90\% larger than 0.4, see fig.~\ref{fig3}(b)).
The inset in fig.~\ref{fig3}(b) shows the relationship of user's reputation and his error
magnitude. The honest user's reputation is decreasing with his error magnitude. When the ratio of spammers is very
large, the decreasing line has larger fluctuation, but the magnitude of fluctuation is very small even when the ratio of spammers is 0.9. 
This shows that the reputations of honest users is decreasing with their error magnitudes.

As discussed above, the correlation based ranking algorithm always gives lower reputations to the random rating spammers, which decreases the 
influence of spammers on the the ranking result. At the same time, the reputations of honest users do not decrease significantly with the increase of
spammers. It implies that the correlation based ranking algorithm can nearly remove the influence of spammers regardless of the ratio of spammers, and
have a high robustness against the attack of random rating spammers.

\subsection{Push rating spamming}

\begin{figure}[h!]
\centering
\scalebox{0.5}[0.5]{\includegraphics{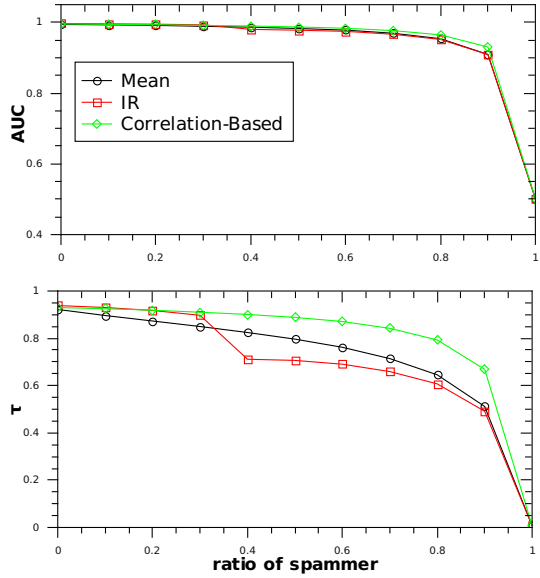} }\caption{
The effectiveness of different algorithms to push rating spamming. The result is obtained by averaging over 10 independently run.}
\label{fig4}
\end{figure}

The effectiveness of different algorithms with spammers who give push ratings is shown in Fig.~\ref{fig4}.
The AUC value of the correlation based ranking algorithm is only slightly higher than the other two algorithms when the ratio of spammers is high, 
but the value of $\tau$ for the correlation based ranking algorithm is significantly higher than the other two algorithm.

As push rating spammers are selected randomly, and every object has the same opportunity to
get push ratings from the spammers, the result is that all object qualities calculated by IR or mean algorithm are higher than expected.
The simulation results imply that, this impact has a great influence on the value of $\tau$ but small influence on the AUC value. A possible reason is that 
the ranking results of IR and mean algorithm have many local fluctuation comparing with the real ranking, and these local oscillations do not influence the AUC value.
As the spammer always gives push ratings, its correlation coefficients with other vectors are always 0. The correlation based ranking algorithm can absolutely remove the
influence of this kind of spammers. So the correlation based ranking algorithm has the highest robustness as indicated by either $\tau$ or AUC.

From the result discussed above we can conclude that, although the IR algorithm has the largest effectiveness for a clean system without spammer,
it is clear that the correlation based algorithm has a good capability to resist spammer attack.

\section{Real data experiment}
After analyses with the artificial data, some real systems are studied in this
section. We use two distinct real data sets containing movie ratings: Netflix
and MovieLens. Movielens is provided by GroupLens project at University of Minnesota
(www.grouplens.org). We use their 1 million ratings data set given on the integer rating scale from 1 to 5.
Each user in Movielens data set has at least 20 ratings. Netflix is huge data set released by the DVD
rental company Netflix for its Netflix Prize
(www.netflixprize.com). We extracted a smaller data set by choosing 4968 users who have rated at
least 20 movies (just like Movielens) and took all movies they had rated. The Netflix ratings are also given on the integer
rating scale from 1 to 5. The characteristics of these data set are summarized in table~\ref{tab2}.

\renewcommand\arraystretch{1.6}
\begin{table}
\caption{Properties of the applied data sets. $|U|$ is the number of users, $|O|$ is the number of objects, $\overline{k_U}$ is
the mean degree of users, $\overline{k_O}$ is the mean degree of objects, and sparsity is the sparsity of the data set. }
\label{tab2}
\setlength\tabcolsep{7.5pt}
\begin{tabular}{cccccc}

\hline
Data set & $|U|$ & $|O|$ & $\overline{k_U}$ & $\overline{k_O}$ & Sparsity\\
\hline	
Movielens & 6040 & 3883 & 166 & 270 & 0.0426\\	
Netflix & 4968 & 16331 & 242 & 74 & 0.0148\\		
\hline
\end{tabular}
\end{table}

As already explained above, one needs an
independently selected set of so-called benchmark objects to test a ranking algorithm on real data. In our tests, we use movies
nominated for the best picture category at the Annual Academy Awards, popularly known as
Oscars (as a source of information we used www.filmsite.org), as benchmark objects.
There are 203 benchmark movies in Movelens data set and 299 in Netflix data set.

The AUC values of different algorithm on real data are shown in table~\ref{tab3}. For the Movielens data set,
IR algorithm has the best effectiveness. While for the Netflix data set, the correlation based algorithm has the best
performance. It is obviously that the AUC values for Movielens using all the three algorithms
are obviously higher than that of Netflix (range from 0.8723 to 0.8763 for Movielens, 0.7609 to 0.7742
for Netflix), and this may suggest that the Netflix data set includes more spammers than
the Movielens data set. Thus based on the results of artificial data, it is suggested that the correlation based ranking algorithm 
obtain better result for Netflix than IR algorithm just because the correlation based ranking algorithm is more robust against spammer attack than IR algorithm.

\renewcommand\arraystretch{1.6}
\begin{table}
\caption{AUC values of different algorithms
for the real data sets. }
\label{tab3}
\setlength\tabcolsep{9pt}
\begin{tabular}{cccc}

\hline
Algorithm & Mean & IR & Correlation-Based \\
\hline		
Movielens & 0.8730 & 0.8763 & 0.8723 \\	
Netflix & 0.7609 & 0.7650 & 0.7742 \\		
\hline
\end{tabular}
\end{table}

\section{Conclusion and discussion}

It is a big challenge to get the right ranking of objects in such rating systems where user vote objects with rating scores, 
especially when spammers are present in the rating system. When it comes to the
user reputation system, how to decide a user's trust value is a crucial question. As correlation is a good way to describe the similarity between two vectors,
we choose correlation coefficient to represent user's reputation and use iterative method to obtain the result step by step. According to
the artificially generated data, the presented correlation based ranking algorithm has a good effectiveness to resist the attack of spammers.
In testing with real data, the present algorithm has a higher
effectiveness than IR algorithm for Netflix, but lower effectiveness for Movielens. That may suggest that Netflix data set includes more spammers than Movielens, and 
the present algorithm has higher robustness to spammers' attack than the other two algorithms.

A good ranking algorithm should be both robust and accurate. The correlation based algorithm presented in this paper
can more effectively tackle the problem of robustness than the others. For the accuracy, there is still
a large room for improvement. On the other hand, how to judge the ranking result is also a problem. For movies, some of them which have not received any award are also widely loved by
people. Only using movies that have been nominated by famous award as benchmark is also not reasonable. The effectiveness of ranking algorithm with artificial data is
easy to evaluate. If real data are completely replaced by artificial data, it will be easier to evaluate a given ranking algorithm.
Our future work will focus on how to build more reasonable models to generate artificial data and improve the accuracy of ranking algorithm.

\acknowledgments
We thank Bill Yeung's help to polish this paper. This work is partially supported by the Swiss National Science Foundation (Project No. 200020-121848).

\end{document}